%% file: paper.tex
\documentclass[useAMS,usenatbib,usegraphicx]{mn2e}
\usepackage{amsmath,amsfonts,amssymb}
\usepackage[latin1]{inputenc}
\usepackage{times} 
\usepackage{color}     
\usepackage{graphicx}
\usepackage{caption}  
\usepackage{subcaption}
\usepackage{float}

\def\Msun{\hbox{$\rm\, M_{\odot}$}}

\newcommand{\spose}[1]{{\hbox to 0pt{#1\hss}}}
\newcommand{\lta}{\mathrel{\spose{\lower 3pt\hbox{$\mathchar"218$}}
     \raise 2.0pt\hbox{$\mathchar"13C$}}}
\newcommand{\gta}{\mathrel{\spose{\lower 3pt\hbox{$\mathchar"218$}}
     \raise 2.0pt\hbox{$\mathchar"13E$}}}

\newcommand{\HI} {H{\sc i} } 
\newcommand{\HIMF} {H{\sc i}MF } 
\newcommand{\HII} {H$_2$ }

\newcommand{\App}[1]{Appendix~\ref{app:#1}}
\newcommand{\Eq}[1]{Equation~\ref{eq:#1}}
\newcommand{\Fig}[1]{Figure~\ref{fig:#1}}
\newcommand{\Sec}[1]{Section~\ref{sec:#1}}
\newcommand{\Tab}[1]{Table~\ref{tab:#1}}

\title [The H{\sc i} mass function of galaxies]{A consistent model for both the
  H{\sc i} and stellar mass functions of galaxies}

\author[Martindale et al.]
{Hazel Martindale$^{1}$\thanks{E-mail: H.Martindale@sussex.ac.uk},
 Peter A. Thomas$^1$, Bruno M. Henriques$^{2,3}$, Jon Loveday$^1$\\
 {}$^1$Astronomy Centre, University of Sussex, Falmer, Brighton BN1 9QH,
 United Kingdom\\
 {}$^2$Max-Planck-Institut f\"ur Astrophysik,
 Karl-Schwarzschild-Str. 1, 85741 Garching b. M\"unchen, Germany\\
 {}$^3$Institute for Astronomy, ETH Zurich, CH-8093 Zurich, Switzerland
}
\begin{document}
\date{Submitted to MNRAS -- \today}

\pagerange{\pageref{firstpage}--\pageref{lastpage}} \pubyear{2016}

\maketitle

\label{firstpage}

\begin{abstract}
Using the L-Galaxies semi-analytic model we simultaneously fit the \HI mass
function, stellar mass function and galaxy colours.  We find good fits to all
three observations at $z = 0$ and to the stellar mass function and galaxy colours
at $z = 2$.

Using Markov Chain Monte Carlo (MCMC) techniques we adjust the L-Galaxies
parameters to best fit the constraining data.  In order to fit the \HI mass
function we must greatly reduce the gas surface density threshold for star
formation, thus lowering the number of low \HI mass galaxies.  A simultaneous
reduction in the star formation efficiency prevents the over production of
stellar content.  A {\it simplified} model in which the surface density
threshold is eliminated altogether also provides a good fit to the data.
Unfortunately, these changes weaken the fit to the Kennicutt-Schmidt relation
and raise the star-formation rate density at recent times, suggesting that a
change to the model is required to prevent accumulation of gas onto dwarf
galaxies in the local universe.

\end{abstract}

\begin{keywords}
methods: numerical -- galaxies: formation -- galaxies: evolution
\end{keywords}

\input{intro.tex}
\input{method.tex}
\input{results.tex}
\input{conc.tex}

\section*{Acknowledgements}
All authors contributed to the development of the model, the interpretation of the results, and the writing of the paper.  HM led the bulk of the data analysis and wrote the initial draft the paper. 

The analysis was undertaken on the {\sc Apollo} cluster at Sussex.
This research made use of Astropy, a community-developed core Python package for Astronomy \citep{Robitaille2013}.
HM(ORCID 0000-0001-6953-7760) acknowledges the support of her PhD studentship from the
Science and Technology Facilities Council (ST/K502364/1). PAT(ORCID 0000-0001-6888-6483) and JL acknowledge support from the Science and Technology Facilities
Council (grant number ST/L000652/1). The work of BH(ORCID  0000-0002-1392-489X) was supported by Advanced Grant 246797 "GALFORMOD" from the European Research Council and by a Zwicky Fellowship.

\bibliographystyle{mn2e}
\bibliography{mn-jour,himass}

\appendix

\input appendix1

\label{lastpage}

\end{document}

%% file: intro.tex
\section{Introduction}
\label{sec:intro}

Cold gas provides the fuel for star formation and understanding its 
properties in galaxies is fundamental to a complete model of galaxy formation. 
While the physics governing the collapse of gas clouds on sub-pc scales, and its 
subsequent conversion into stars, remain largely unknown, simulations can be 
used to explore the factors that affect the gas and ultimately the stellar content 
of galaxies.

The relations governing star formation link the cold gas content to the amount
of stars formed.  The widely used Kennicutt-Schmidt
relation~\citep{Schmidt1959,Kennicutt1998} is a relation between total cold gas
content and the star formation rate of a galaxy.  More recent observations,
however, have shown the correlation to be stronger with only the molecular, \HII
component of cold gas \citep{Bigiel2008,Leroy2008}. 

\HII gas is not directly detected and is instead observed via the tracer
molecule CO which adds uncertainty to these measurements. The HI 
component, on the other hand, correlates more weakly with star formation than 
the \HII, but can be directly observed through the 21 cm emission.
  \HI surveys such as the \HI Parkes ALL-Sky
Survey (HIPASS; \citealt{Meyer2004}) and the Arecibo Legacy Fast ALFA survey
(ALFALFA; \citealt{Giovanelli2005}) now provide large samples of statistical
significance. The \HI mass function from these surveys measures masses down to
$10^6 \mathrm{M_{\odot}}$ allowing galaxy gas content to be probed across a full
range of masses \citep{Zwaan2005,Martin2010}.  Up coming surveys at new
facilities such as the Australian SKA pathfinder (ASKAP,
\citealt{Johnston2008}), Karoo Array Telescope (MeerKAT, \citealt{Booth2009})
and the Square Kilometre Array (SKA\footnote{\tt
  https://www.skatelescope.org/project/}) will greatly improve the observational
constraints on \HI content of galaxies. For that reason, we choose to use
\HI as a constraint in our models.

Semi-analytic models (SAMs) provide a framework to explore the statistical
properties of the observed galaxy population. The evolution of large scale 
structures is given by dark matter merger trees, either from N-body simulations or 
analytic calculations, and the baryonic component is modelled via empirical relations 
that are designed to capture the key physics \citep{White1988,Cole1991,Lacey1991,White1991,Kauffmann1993,Kauffmann1999,Somerville1999,Springel2001,Hatton2003,Kang2005,Croton2006, DeLucia2007,Guo2011,Lu2011,Benson2012}. 
A downside of SAMs is that they necessarily impose restrictive assumptions about the 
geometry of galaxies and the exchange of material with their surroundings. The major 
advantage over hydrodynamical simulations is that they are quick to run allowing us 
to explore the impact of different implementations of physical processes and 
different parameter values. In recent years, the introduction of robust statistical 
methods has even allowed the full exploration of parameter space
\citep{Kampakoglou2008,Henriques2009,Benson2010,Bower2010,Henriques2010,Lu2011,Lu2012,Mutch2013,Henriques2013,Benson2014,Ruiz2015}.

The most recent version of the L-Galaxies SAM \citep[hereafter HWT15]{HWT15}
provides an excellent fit to a wide range of galaxy properties across a wide
range of redshifts. 
In this paper we aim to improve the agreement between the HWT15 model 
to the HI mass function by including it as an extra constraint in addition to the stellar mass
function and galaxy colours.  We find that we can obtain a good fit to all
data-sets simultaneously by lowering, or even eliminating altogether, the
surface density threshold for star formation. Unfortunately, these changes
weaken the fit to the Kennicutt-Schmidt relation and raise the
star-formation-rate density at recent times, suggesting that a change in the
model is required to prevent accumulation of gas onto dwarf galaxies in the
local Universe.

The paper is structured as follows: In \Sec{method} we describe the L-Galaxies
semi-analytic model and the method of gas division.  In \Sec{results} we present
the results of constraining the model with the HI mass function in addition to
the galaxy colours and stellar mass function.  In \Sec{discuss} we examine which
parameters have changed in order to produce a good fit to all constraining data
sets and compare our results to the Kennicutt-Schmidt relation.  We provide our
conclusions in \Sec{conclusions}.

%% file: method.tex
\section{Method}
\label{sec:method}

\subsection{L-Galaxies}
Semi-analytic models provide a tool to explore galaxy formation and evolution
and simulate the cosmic galaxy population.  The models use coupled differential 
equations to follow the evolution of the baryonic component of galaxies 
usually constructed on top of dark matter halos from an $N$-body
simulation.  Many aspects of galaxy formation are included in these models such
as, star formation, gas cooling, metal enrichment, black hole growth and
feedback processes.

The Munich SAM, L-Galaxies, has been developed over many years using galaxy
formation recipes to match the observed galaxy populations
\citep{White1988,Kauffmann1993,Kauffmann1999,Springel2001,Springel2005,Croton2006,DeLucia2007,Guo2011,Guo2012,Henriques2013,HWT15}.  The underlying merger trees are
extracted from the Millennium \citep{Springel2005} and MillenniumII Simulations
\citep{Boylan2009}.  The latest version of the model, on which this work is
based, is given in HWT15.  This version uses {\it Planck} year 1 cosmology with
the Millennium dark matter merger trees scaled according to the method of
\citet{AnW10} \citep[as updated by][]{AnH15}.  HWT15 constrain the model to give
a good fit to the stellar mass function and the galaxy colours over the redshift
range 0-3.  A full description of the model is given in the supplementary
material of HWT15.

\subsubsection{MCMC}
Having many recipes controlling galaxy formation gives rise to numerous free
parameters which, when considering individual galaxy properties independently,
are frequently degenerate with each other.  It would be a long
and inefficient process of trial and error to adjust the parameters to best fit
the observations by hand when alterations to the model are made.  We employ the
MCMC procedure within L-Galaxies to find a best fit set of parameters
\citep{Henriques2009,Henriques2013}.  This method approximates a likelihood
value for the ability of the model to recover the observed galaxy property and
then uses the MCMC technique to minimise that value and locate a best set of
parameters. 

\subsubsection{Star formation law}
\label{sec:sfLaw}
In the model we assume stars form from the total cold gas within a given
galaxy's disk (i.e.~the model does not distinguish between \HI\ and molecular
gas).  The star formation rate is given by
\begin{equation} \label{eq:SFLaw}
\dot{M}_{\mathrm{stellar}} = \alpha_\mathrm{SF} \frac{\left(M_{\mathrm{gas}} -
  M_{\mathrm{crit}}\right)}{t_{\mathrm{dyn,disk}}},
\end{equation}
where $\alpha_\mathrm{SF}$ is a normalisation parameter, $M_{\mathrm{gas}}$ is
the total cold gas mass, $t_{\mathrm{dyn,disk}}$ is the dynamical time, and
$M_{\mathrm{crit}}$ is a threshold mass whose need is based on a long-standing
acceptance that there is a minimum surface density required for star formation
\citep{Kauffmann1996,Kennicutt1998}.  Based on the argument in
\citet{Kauffmann1996} we take $M_{\mathrm{crit}}$ to have the form
\begin{equation}\label{eq:MCrit}
M_{\mathrm{crit}}=M_{\mathrm{crit,0}}
\left(V_\mathrm{200c}\over200\,\mathrm{km}\,\mathrm{s}^{-1}\right)
\left(R_\mathrm{gas}\over 10\,\mathrm{kpc}\right),
\end{equation}
where $V_\mathrm{200c}$ is the virial speed of the halo, $R_\mathrm{gas}$ is the
gas disk scale-length, and $M_{\mathrm{crit,0}}$ is a normalisation constant.
Since Kauffmann et al. (1999) and prior to HWT15, all versions of the Munich
model fixed $M_{\mathrm{crit,0}}=3.8\times10^9\,\Msun$.  Recent work indicates
that star formation is linked more closely to the molecular gas than to the total
gas content \citep{Bigiel2008,Leroy2008}.  This allows stars to form in regions
with smaller total gas thresholds than previously and to allow for this we now
(from HWT15 onwards) treat $M_{\mathrm{crit,0}}$ as a free parameter.

\begin{figure}
\centering
\includegraphics[width=8.6cm]{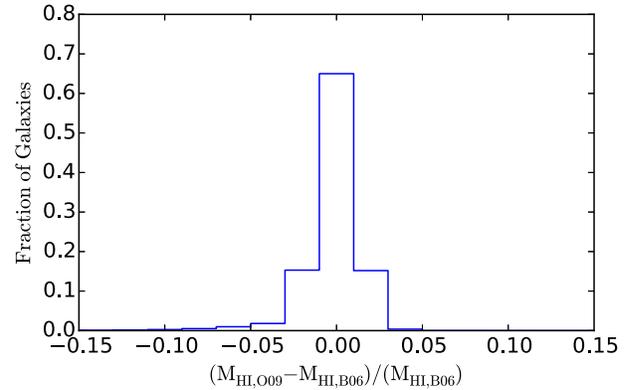}
\caption{Histogram showing the difference between the HI mass
  calculated using BR06 and the approximation of O09. }
\label{fig:approx_hist}
\end{figure}

\subsection{The \HI model}
\label{sec:HImodel}

We use the model of \citet[hereafter BR06]{Blitz2006} to divide the cold gas
into its \HI and \HII components in post processing.  In this model the ratio of
\HI to \HII gas in a galaxy is determined by mid-plane hydrostatic pressure in
the galactic disk.  \citet{Elmegreen1989,Elmegreen1993} propose a form for the
mid-plane pressure
\begin{equation} 
  \label{eq:pressure}
 P_{\mathrm{ext}} \approx \frac{\pi}{2}
 G\Sigma_{\mathrm{gas}}\left(\Sigma_{\mathrm{gas}}+\Sigma_{\mathrm{star}}
   \frac{c_{\mathrm{gas}}}{c_{\mathrm{star}}}\right),
\end{equation}
where $\mathrm{\Sigma_{\mathrm{gas}}}$, $\mathrm{\Sigma_{\mathrm{star}}}$ are
the cold gas and stellar surface densities, $\mathrm{c_{\mathrm{gas}}}$,
$\mathrm{c_{\mathrm{star}}}$ are the gas and stellar vertical velocity
dispersions and G is the gravitational constant.  The mid-plane pressure is
calculated from the equations of hydrostatic equilibrium for a thin disk of gas
and stars.
This pressure is an important factor in the formation of giant
clouds within which \HII is found.  BR06 make the assumption that the
ratio of \HII to \HI in the galaxy is a function of the pressure given
in~\eqref{eq:pressure}.  The relation takes the form of a power-law:
\begin{equation}
  \label{eq:rmol_p}
  R_{\mathrm{mol}}=\frac{\Sigma_{\mathrm{H2}}}{\Sigma_{\mathrm{HI}}}=\left(\frac{P_{\mathrm{ext}}}{P_0}\right)^{\alpha}
\end{equation}
where $\mathrm{\Sigma_{H2}}$ and $\mathrm{\Sigma_{HI}}$ are the disk surface
densities of \HII and \HI gas respectively and $\mathrm{P_0}$ and
$\mathrm{\alpha}$ are fitting constants.  This was further explored using
resolved observations of galaxies \citep{Blitz2006,Leroy2008}.

This model of gas division requires information on the radial distribution of gas 
inside galaxies.  In order to include it at each step of the L-Galaxies MCMC chain
without prohibitively slowing the calculation we use the approximation to BR06
model derived in \citet[hereafter O09]{Obreschkow2009}.  They write
$R_{\mathrm{mol}}$ as,
\begin{equation}
R_{\mathrm{mol}} = R_{\mathrm{mol}}^c \exp(-1.6\,r/r_{\mathrm{disk}}),
\end{equation}
where $r_{\mathrm{disk}}$ is the scale length of the gas disk and
$R_{\mathrm{mol}}^c$ is
\begin{equation}
R_{\mathrm{mol}}^c = \left[K r_{\mathrm{disk}}^{-4} M_{\mathrm{gas}}(M_{\mathrm{gas}} + \left<f_{\sigma}\right>M_{\mathrm{disk}}^{\mathrm{stars}}) \right]^{\alpha},
\end{equation}
where $M_{\mathrm{gas}}$ is the total cold gas mass,
$M_{\mathrm{disk}}^{\mathrm{stars}}$ is the mass of the stellar disk and $K =
G/(8 \pi P_0)$.  We adopt the same values of constants as O09: $P_0 = 2.34
\times 10^{-13} Pa$, $\alpha = 0.8$ and $\left<f_{\sigma}\right> = 0.4$.
Through $R_{\mathrm{mol}}$ we can derive expressions for the surface density of
\HI and \HII which when integrated give the M$_{\mathrm{HI}}$ and
M$_{\mathrm{H2}}$.

O09 approximate the integration, finding that the ratio of \HII to \HI is given
by
\begin{align} \label{eq:approx}
\frac{M_{\mathrm{H2}}}{M_{\mathrm{HI}}} &= \frac{\int \Sigma_{\rm H2}(r)~ \mathrm{d}A}{\int \Sigma_{\rm \mathrm{HI}}(r)~\mathrm{d}A}\nonumber \\ 
 & \approx \left(3.44 R_{\mathrm{mol}} ^{c \ -0.506} + 4.82 R_{\mathrm{mol}}^{c \ -1.054}  \right)^{-1}. 
\end{align}
Using this approximation along with assuming that $M_{\mathrm{H}} =
M_{\mathrm{HI}}+M_{\mathrm{HII}}$ we can calculate the masses without dividing
the galaxies into rings and significantly speed up the calculation.  We assume
that $M_{\mathrm{H}} = 0.74 M_{\mathrm{cold gas}}$.  In \Fig{approx_hist} we
test the accuracy of this approximation for galaxies produced by our model.  We
show the difference in mass calculated using both approaches and find the
agreement between the two methods to be excellent.  We
agree with the statement of O09 that the accuracy is greater than 5 per cent.

\subsection{Observational Constraints}
\label{sec:ObsConst}
We constrain the model using observations at $z=0$ and $z=2$. 
At $z=0$ we use: 
\begin{itemize}
\item The stellar mass function is a combination of the SDSS \citep{Li2009} and GAMA \citep{Baldry2012} results.
\item The \HI mass function is from HIPASS \citep{Zwaan2005}.
\item The red fraction is obtained by dividing the stellar mass function of red galaxies by the sum of the red and blue stellar mass functions. We use data from \cite{Bell2003} and \cite{Baldry2012}. 
\end{itemize}

At $z=2$:
\begin{itemize}
\item The stellar mass function is a combination of COSMOS \citep{Sanchez2011}, ULTRAVISTA \citep{Ilbert2013,Muzzin2013} and ZFOURGE \citep{Tomczak2014}.
\item The red fraction of galaxies also uses COSMOS \citep{Sanchez2011}, ULTRAVISTA \citep{Ilbert2013,Muzzin2013} and ZFOURGE \citep{Tomczak2014}.
\end{itemize}

%% file: results.tex
\section{Results}
\label{sec:results}
We present results for several different versions of the model:
\begin{itemize}
\item HWT15 (green dash-dotted line): The reference model, which did not use the
  \HIMF as a constraint.
\item HIConstraint (red solid line): The HWT15 model but adding in the \HIMF as a
  constraint at $z=0$.
\item NoSFThreshold (blue dashed line): The same as the HIConstraint but with
  the minimum threshold surface density for star formation set equal to zero.
\item DLB07Reincorporation (yellow dotted line): As for the HIConstraint but
  using the older \cite[hereafter DLB07]{DeLucia2007} recipe for reincorporation
  of ejected material.
\end{itemize}
All of the the models were constrained to simultaneously match the observations
described in \Sec{ObsConst}, except HWT15 which did not use the \HIMF as a
constraint.

\subsection{\HI Mass Function}

\begin{figure}
\includegraphics[width=8.6cm]{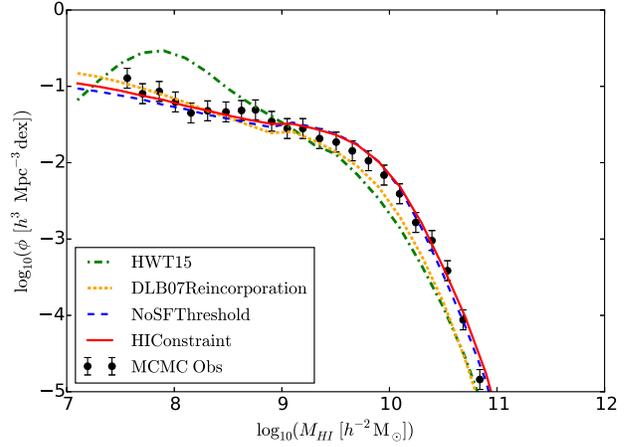}
\caption{The \HI mass function at z=0. The black points are the observed \HI
  mass function from HIPASS. The coloured lines represent the different models:
  green, HWT15; red, HIConstraint; blue, NoSFThreshold; yellow, DLB07
  Reincorporation.}\label{fig:himf}
\end{figure}

\label{sec:HIresult}
The \HI mass function is shown in \Fig{himf}.  It is immediately obvious that
the HWT15 reference model is a poor fit to observations.  This is not an inherent
deficiency of the model, but results from the fact that the observed mass
function was not used as an input constraint.  The HWT15 model does, in fact,
provide a slightly better fit overall to the stellar masses and galaxy colours
at $z=0$ \& 2, than the HIConstraint or NoSFThreshold model, but the difference
is slight.  That goes to show that the \HI mass function serves as a largely
independent constraint.

The HIConstraint model, however, that does use the \HI\ as an additional
constraint, provides a very good fit to the \HI\ mass function.  It does that
largely by reducing the $\Sigma_{\mathrm{SF}}$ parameter in the model that
governs the minimum surface density for quiescent star formation (see
\Tab{parameters}).  This allows more cold gas to be consumed in low-mass
galaxies. In order to maintain the same overall stellar mass, the star formation
efficiency is reduced leading to a reduction of gas consumption in high-mass
galaxies.

Because the HIConstraint model lowers the minimum surface density for star
formation so much, we also examined a NoSFThreshold model in which it is set
equal to zero (thus reducing the number of free parameters in the model by one).
The two are barely distinguishable in their predictions (except that the
NoSFThreshold model has slightly bluer colours -- see \Sec{galcol}).

To try to understand why \citet[hereafter Lu14]{Lu2014} have claimed that it is
not possible to reproduce the \HI\ mass function, we also ran a model that is
identical in every respect to the HIConstraint model, except that the
reincorporation timescale follows the parameterisation given in DLB07 rather
than HWT15.  This DLB07 model, which uses HI as a constraint,  provides a better fit than the original HWT15 but
is clearly a significantly worse than either the HIConstraint or NoSFThreshold
models.  This will be discussed further in \Sec{otherwork} below.

\begin{figure}
\centering
\begin{subfigure}{8.6cm}
\includegraphics[width=8.6cm]{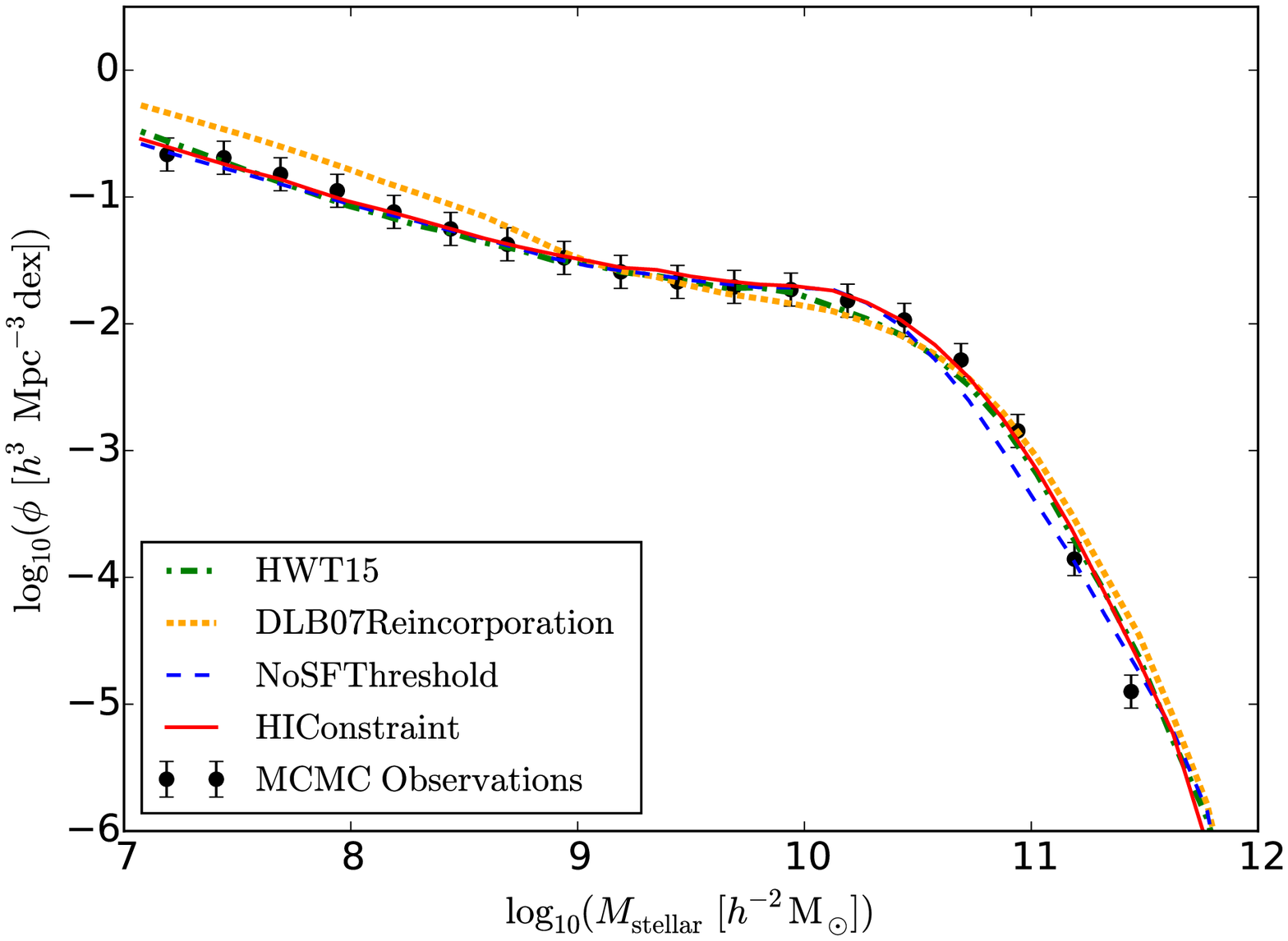}
\end{subfigure}

\begin{subfigure}{8.6cm}
\includegraphics[width=8.6cm]{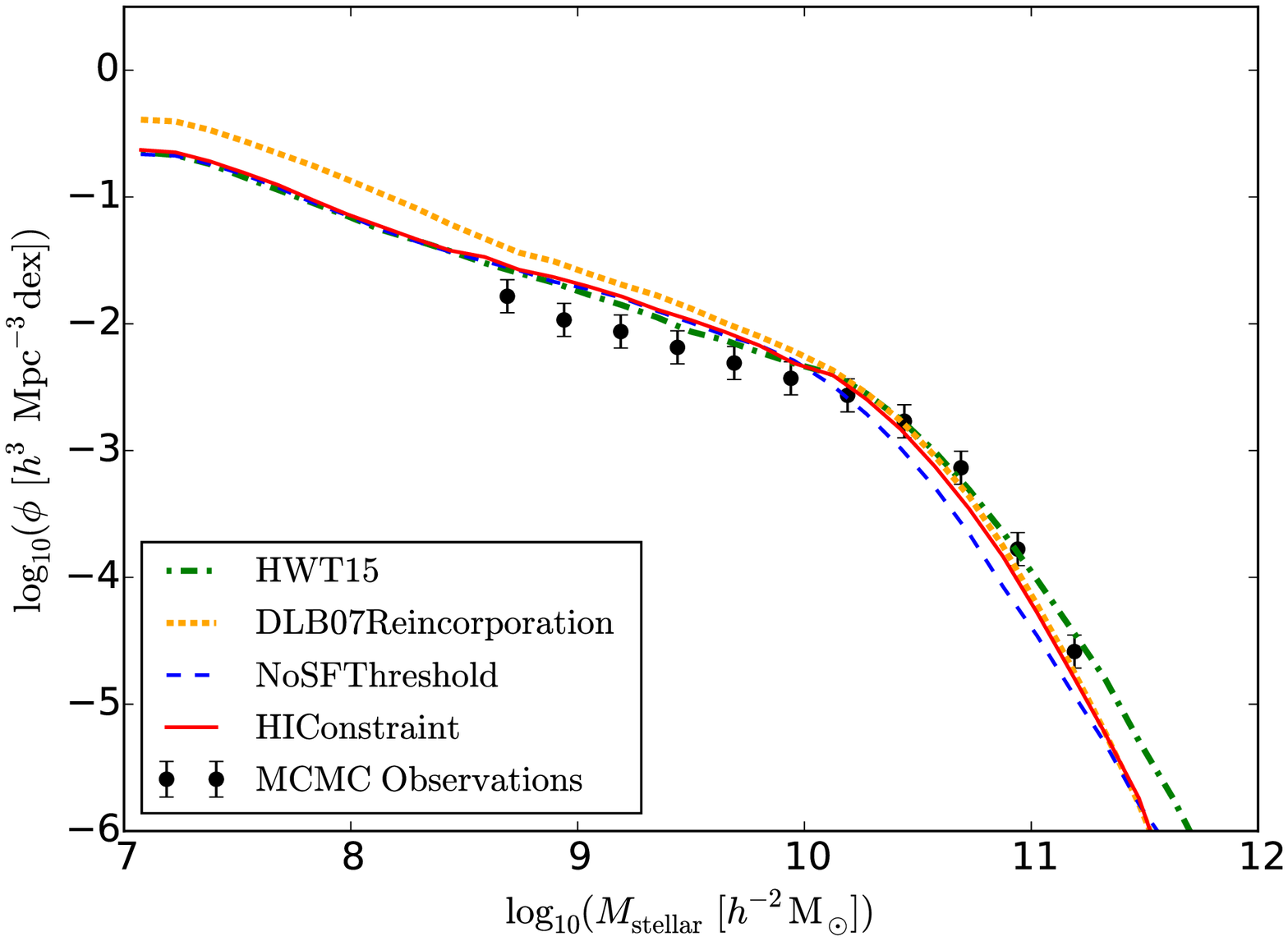}
\end{subfigure}
 \caption{The stellar mass function, $z=0$ is shown in the upper panel and $
   z=2$ is shown in the bottom panel.  The black points are the observations
   used within the MCMC as constraints. The coloured lines are as in
   \Fig{himf}.}
\label{fig:smf}
\end{figure}

\subsection{Stellar Mass Function}
\label{sec:SMresult}
The stellar mass function is shown in \Fig{smf}, the upper panel showing $z=0$
and the lower $z=2$.  At $z=0$ we find an excellent fit to the observed stellar
mass function in both the HIConstraint and NoSFThreshold models, better even
than that of the reference model of HWT15.  There is no significant difference
between the red and blue lines indicating that a non-zero threshold cold gas
surface density is not required to fit the stellar mass function at $z=0$.  The
DLB07 reincorporation model provides a significantly worse fit both at the knee
of the SMF and the slope at low-masses.  This is discussed further in
\Sec{otherwork}.

The fit at $z=2$ is a marginally worse than in HWT15 for both our models
although the difference is very small.
The DLB07 reincorporation model again fares worse than the others.

\begin{figure}
\begin{subfigure}{8.6cm}
\includegraphics[width=8.6cm]{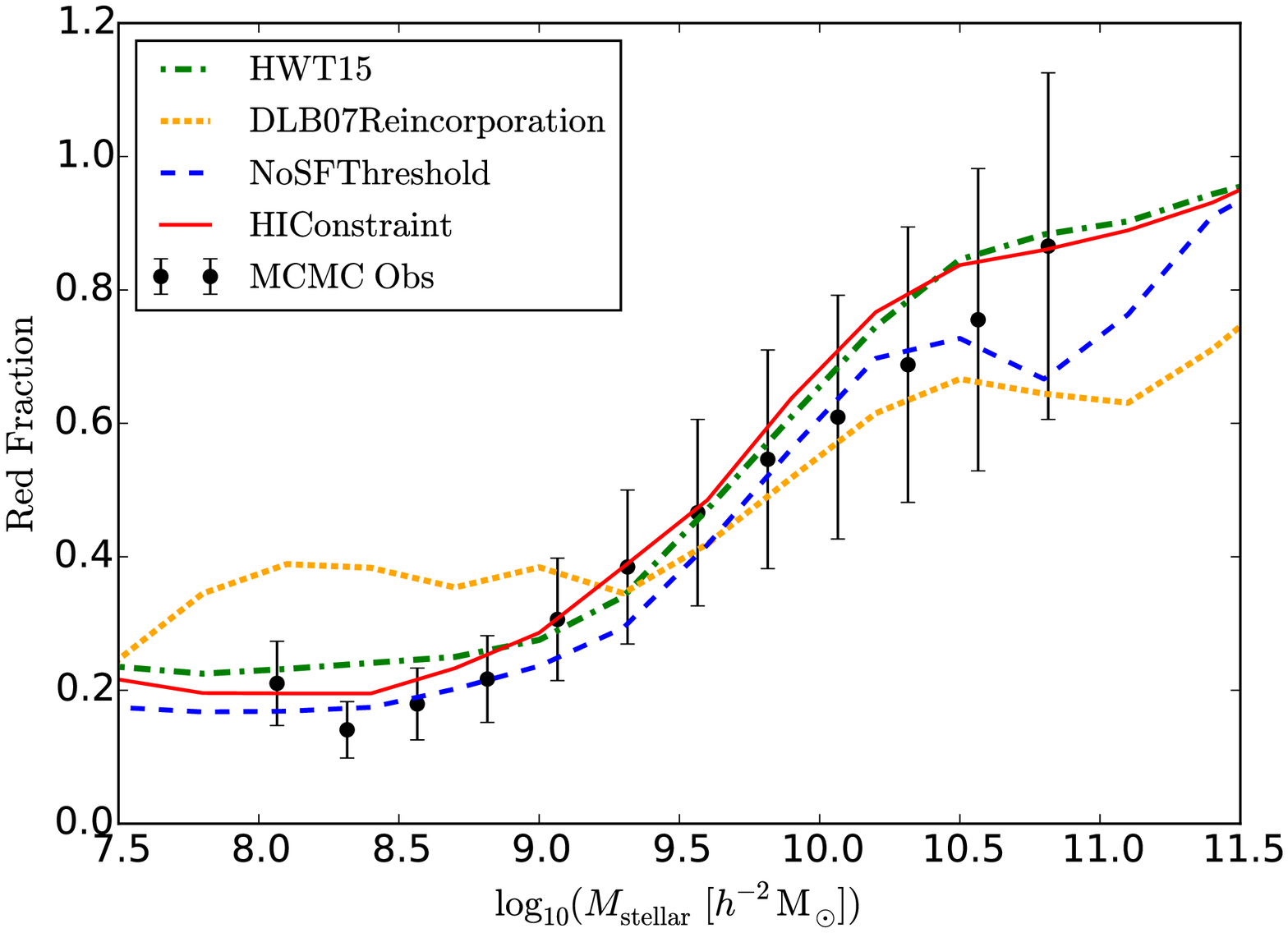}
\end{subfigure}

\begin{subfigure}{8.6cm}
\includegraphics[width=8.6cm]{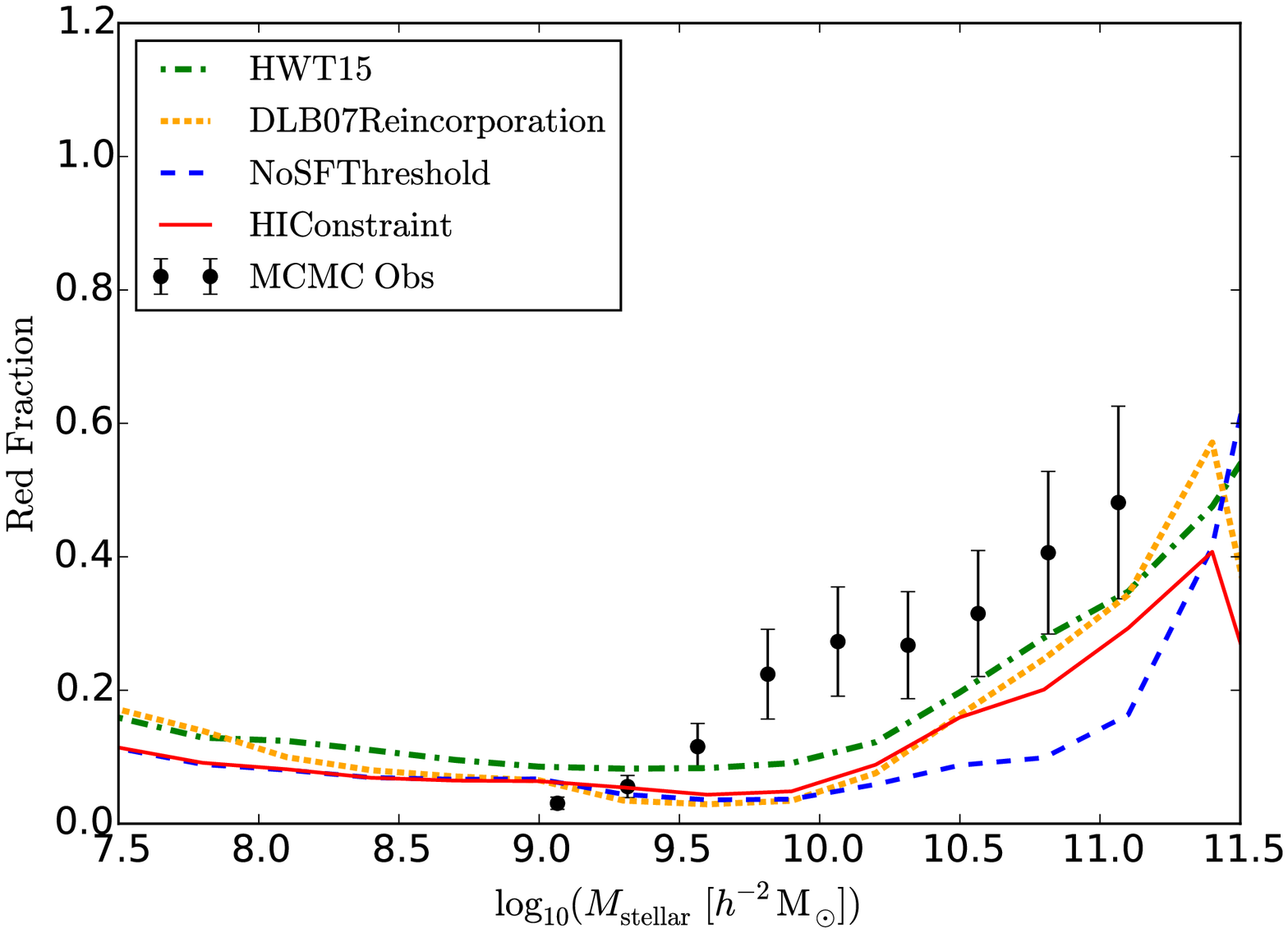}
\end{subfigure}
\caption{The red fraction mass function shown in the upper panel is $z=0$ and
  $z=2$ is shown in the lower.  The line colours refer to the same models as
  those in \Fig{smf}.  The black points are the observed red fractions used with
  in the MCMC. }
\label{fig:redfrac}
\end{figure}

\subsection{Galaxy Colours}
\label{sec:galcol}
The model was also constrained using the red fraction of galaxies, with the
same prescription as HWT15.  The red fraction is shown in \Fig{redfrac} with
$z=0$ in the upper panel and $z=2$ in the lower panel.  At $z=0$ we have similar
fits to HWT15, except for the DLB07 model which has too few red galaxies at high
masses and too many at low masses.

At $z=2$ all models under predict the fraction of red galaxies at high stellar
mass, with the NoSFThreshold model this time giving the poorest fit to the data.
The decrease in the red population at $z=2$ indicates the model has too much
ongoing star formation in the highest mass galaxies. This suggests that the
reduction of the threshold for star formation may not be an ideal solution to
our problem, as discussed further in \Sec{KSlaw}, below.

\begin{figure}
\begin{subfigure}{8.6cm}
\includegraphics[width=8.6cm]{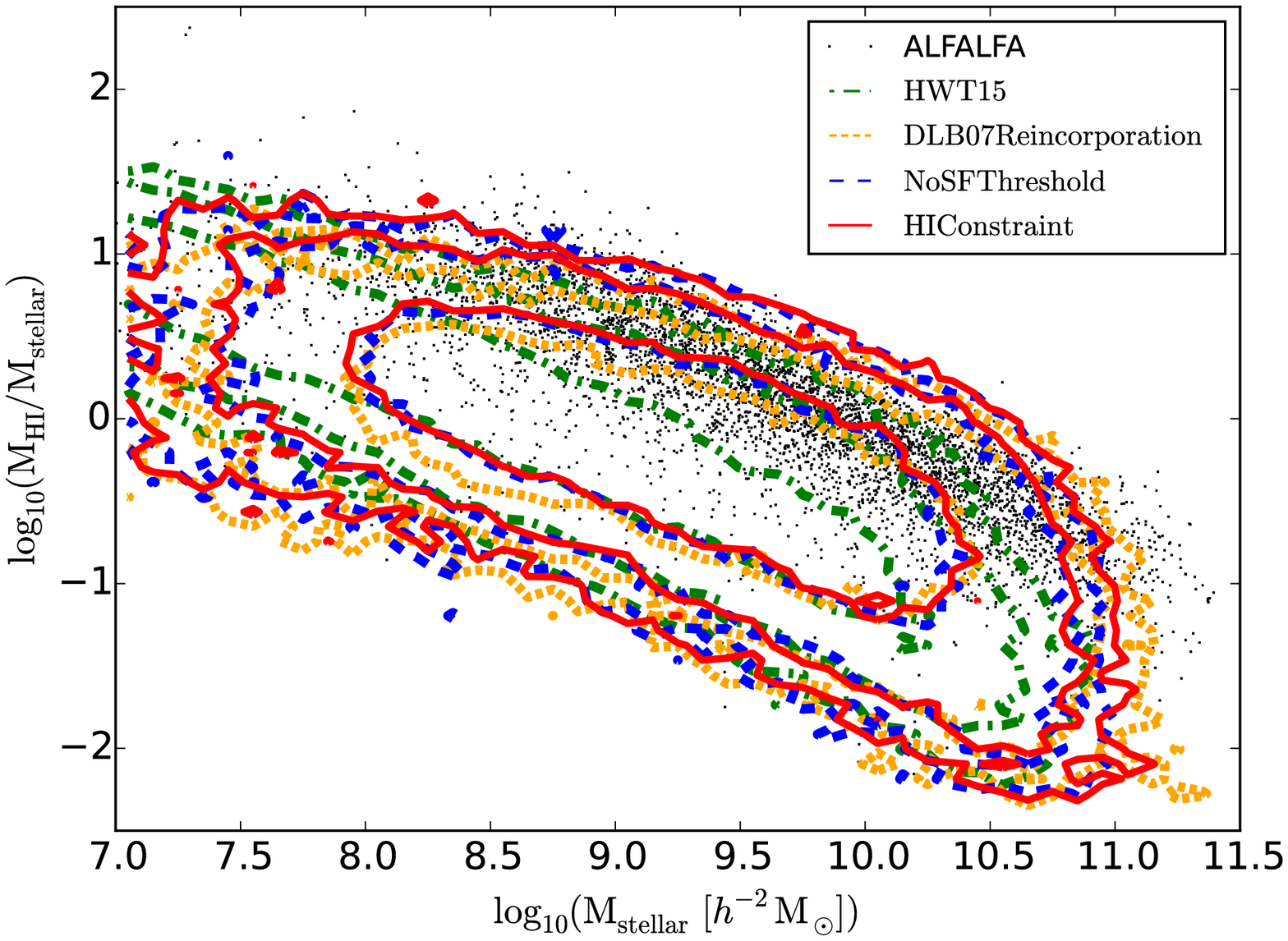}
\end{subfigure}

\begin{subfigure}{8.6cm}
\includegraphics[width=8.6cm]{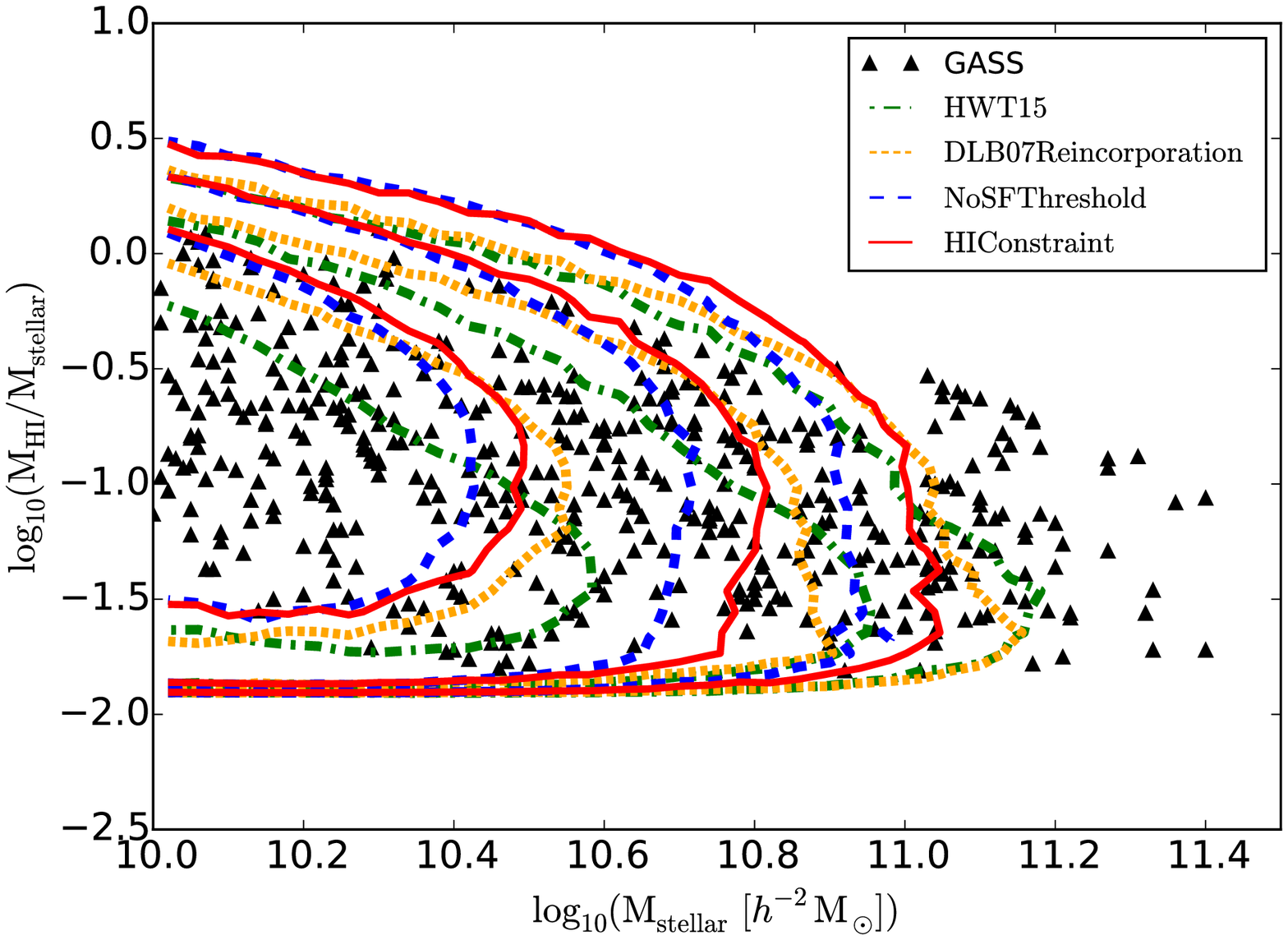}
\end{subfigure}
\caption{The HI to stellar mass fraction. Compare the HI mass to stellar mass
  fraction to observational data shown in black.  The top panel compares our
  data to that from the ALFALFA survey, \citep{Haynes2011}. The lower panel
  compares with the GASS survey, triangles, \citep{Catinella2013}. We show each
  model as coloured contours. The dotted contour encloses 99 per cent of the data, the
  dashed contour encloses 95 per cent and the solid 68 per cent. For each survey we attempt to
  mimic the selection of each survey with the model data before comparing. }
\label{fig:gasfrac}
\end{figure}

\subsection{Gas Fractions}
\label{sec:gasfrac}

We calculate the \HI to stellar mass ratio and compare to those observed by the ALFALFA, \citep{Haynes2011}, and GASS, \citep{Catinella2013}, surveys.
In general we have good agreement with the observed \HI gas fractions shown in \Fig{gasfrac}. The top panel of \Fig{gasfrac} compares the models to ALFALFA while the bottom compares to GASS. The contour levels shown
in \Fig{gasfrac} for each model enclose 68, 90 and 99 per cent of the data. Our models galaxies reproduce the observations of the the GASS survey much more closely than those of ALFALFA. 

The ALFALFA survey is a flux limited survey and due to limited sensitivity it can not observe very low \HI mass galaxies. 
This leads to the survey missing low \HI flux objects with correspondingly low gas fractions. 
In order to perform a detailed comparison we would need to precisely mimic the survey selection of ALFALFA in the model galaxies. 
In this work we perform a crude selection on the semi-analytic galaxies, converting the \HI mass to a \HI flux by setting an observer at the centre of the simulation box. We see from the top panel of \Fig{gasfrac} that our model galaxies span the same range of stellar mass as the ALFALFA data and show the same upper limit in gas fraction \citep{Maddox2015}. 
However, the median ratio is offset significantly from that observed by ALFALFA. 
The comparison with an additional data set in the bottom panel suggests that this might result from the observational selection being more complex than the crude flux cut we have applied to the model galaxies. For example, there is also a dependence on the width of the observed spectral line.

In the lower panel of \Fig{gasfrac} we compare our model results to the GASS survey. 
This should represent a more meaningful comparison since observations are stellar mass selected 
and have a much lower sensitivity to \HI detection. Here, theoretical predictions match the observed
gas fractions reasonably well. 

In both the top and bottom panels of \Fig{gasfrac} all 4 versions of the model are shown to produce similar results. The gas fractions is not currently a good measure to distinguish between the model versions. But, consistent with our findings on the mass functions, the original HWT15 model seems to have a lower \HI mass fraction in high-mass galaxies than do the other models that use \HI as a constraint.

%% file: conc.tex
\begin{table*}
\begin{minipage}{17.2cm}
\centering
  \begin{tabular}{c|c|c|c|c|c}
    
    \hline 
    Parameter & HWT15 & HIConstraint & NoSFThreshold & DLB07 Reincorporation &Units \\
    \hline \hline 
    $\alpha_{\mathrm{SF}}$ (SF eff) & 0.025 & 0.0081 & 0.012 & 0.0084 &\\
    $\mathrm{\Sigma_{SF}}$ (SF gas density threshold )
    & 0.24 & 0.0018 & 1e-6 & 0.0024 & $10^{10}\ M_{\odot}\ pc^{-2}$ \\
    $\alpha_{\mathrm{SF,burst}}$ (SF Burst eff) & 0.60 & 0.92 & 0.68 & 0.54 & \\
    $\beta_{\mathrm{SF,burst}}$ (SF Burst Slope) & 1.9 & 1.7 & 1.4 & 0.86  &\\
    \hline \hline
    $k_{AGN}$ (Radio feedback eff)  & 0.0053 & 0.01 & 0.025 & 7.2
    $\times 10^{-4}$ & $\mathrm{M_{\odot}\ yr^{-1}}$\\
    $f_{BH}$ (Black hole growth eff) & 0.041 & 0.042 & 0.022 & 0.030 & \\
    $V_{BH}$ (Quasar growth scale) & 750 & 900 & 840 &  300 &$\mathrm{km\ s^{-1}}$ \\
    \hline 
    $\epsilon$ (Mass-loading eff) & 2.60 & 1.9 & 1.5 & 3.06 &\\
    $V_{reheat}$ (Mass-loading scale) & 480 & 270 & 370 & 100 & $\mathrm{km\ s^{-1}}$ \\
    $\beta_1$ (Mass-loading slope) & 0.72 & 1.1 & 0.55 & 3.8 &\\
    $\eta$ (SN ejection eff) & 0.62 & 0.18 & 0.36 & 0.22 &\\
    $V_{\mathrm{eject}}$ (SN ejection scale)  & 100 & 200 & 120 & 150 & $\mathrm{km\ s^{-1}}$\\
    $\beta_2$ (SN ejection Slope) & 0.80 & 2.1 & 3.9 & 3.2 &\\
    $\gamma$ (Ejecta reincorporation)  & 3.0 $\times10^{10}$ & 2.2$\times 10^{10}$ &
    2.1$\times10^{10}$& 0.35 & $\mathrm{yr}$\\
    \hline 
    $y$ (Metal yield) & 0.046 & 0.035 & 0.027 & 0.021 &\\
    \hline 
    $R_{\mathrm{merger}}$ (Major-merger thereshold) & 0.10 & 0.43 & 0.37 & 0.33 &\\
    $\alpha_{\mathrm{friction}}$ (Dynamical friction) & 2.5 & 4.5 & 4.3 & 2.5 &\\
    $M_{\mathrm{r.p.}}$ (Ram-pressure threshold) & 1.2$\times 10^{4}$ &2.6 $\times 10^{4}$ &
    2.0 $\times 10^{4}$ & 1.1 &  $\mathrm{10^{10}\ M_{\odot}}$\\
	\hline
  \end{tabular}
\end{minipage}
\caption{Parameters constrained by the MCMC model. Best fit parameters are given
  for each model as well as HWT15 for comparison.}
\label{tab:parameters}
 \end{table*}

\section{Discussion}
\label{sec:discuss}
\subsection{Changes to model parameters} \label{sec:paramchange}

We start our discussion with the original HWT15 model and the new HIConstraint
and NoSFThreshold models.  We defer the discussion of the DLB07 model to the final
paragraph of this section and \Sec{otherwork}.

\begin{figure*}

\begin{minipage}{17.2cm} 
\begin{subfigure}{8.6cm}
\includegraphics[width=8.6cm]{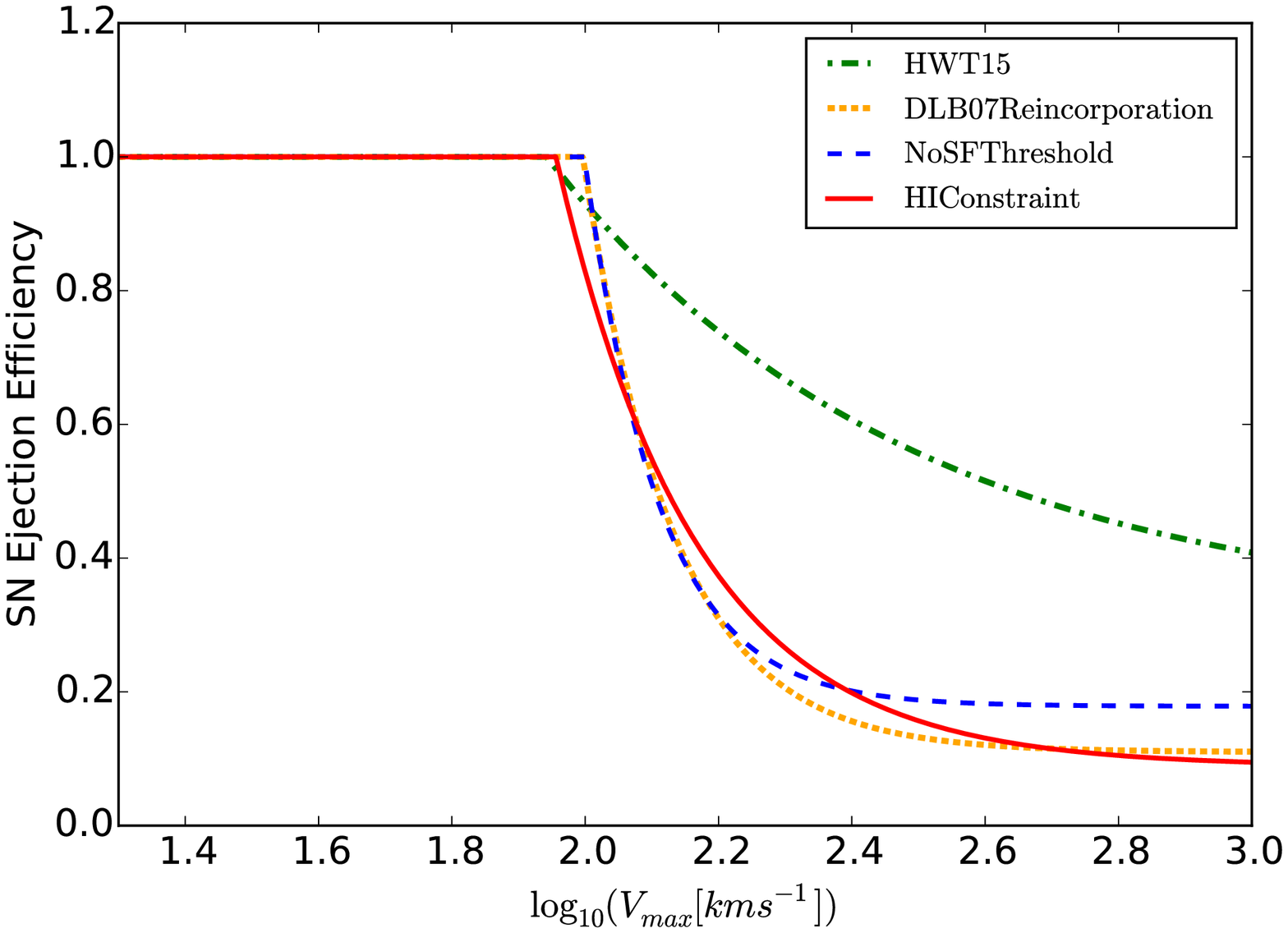}
\end{subfigure}
\begin{subfigure}{8.6cm}
\includegraphics[width=8.6cm]{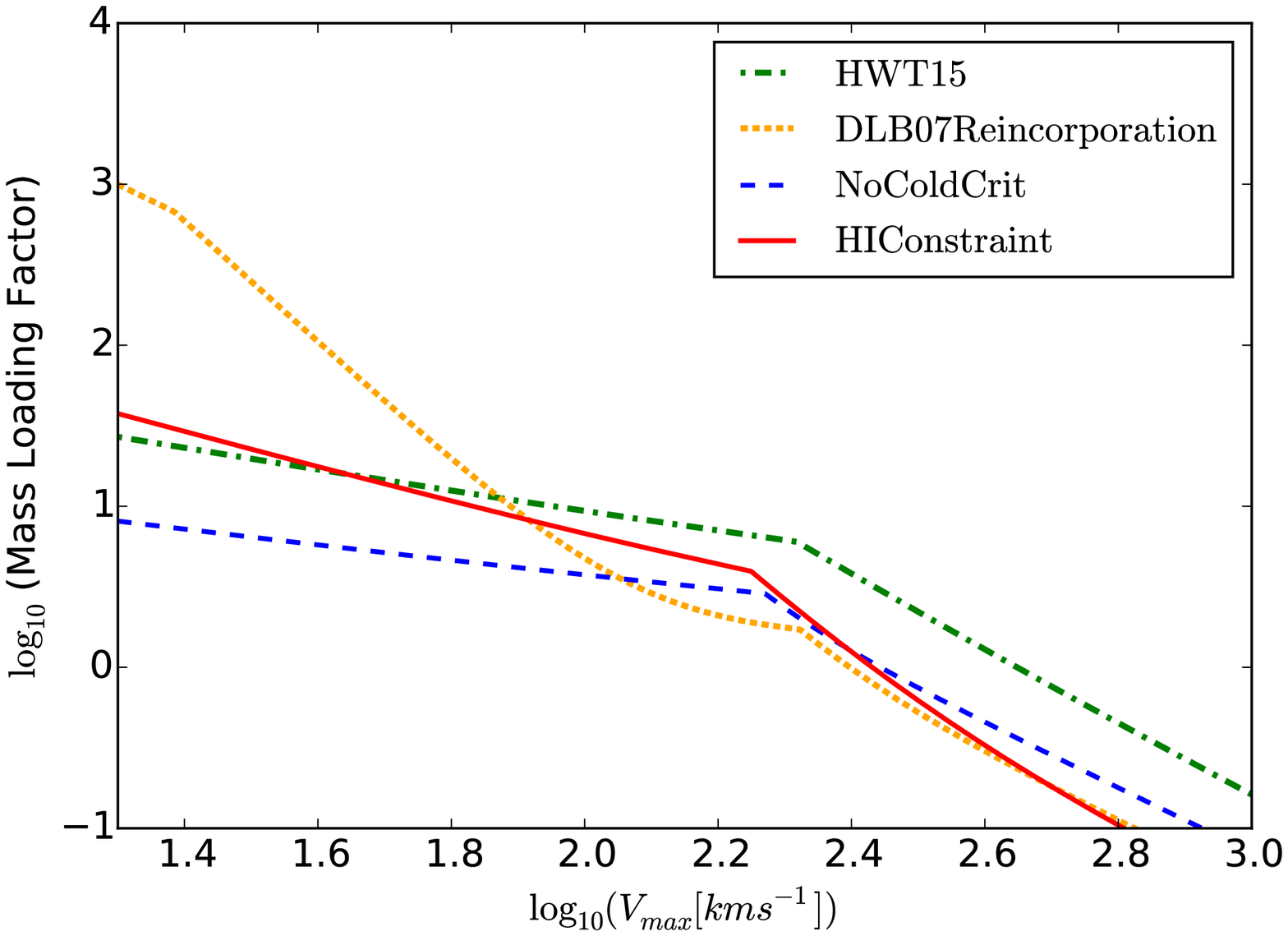}
\end{subfigure}

\begin{subfigure}{8.6cm}
\includegraphics[width=8.6cm]{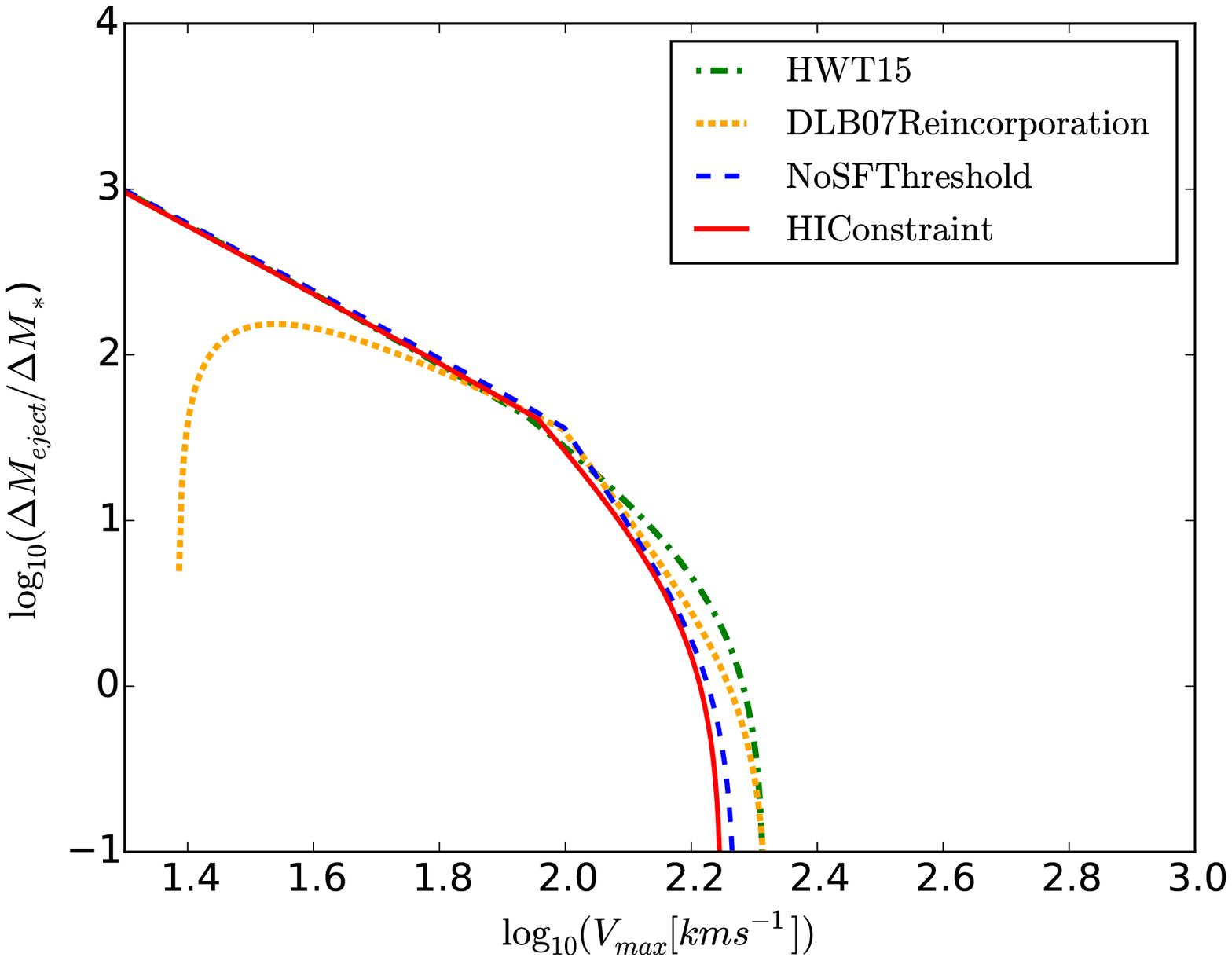}
\end{subfigure}
\begin{subfigure}{8.6cm}
\includegraphics[width=8.6cm]{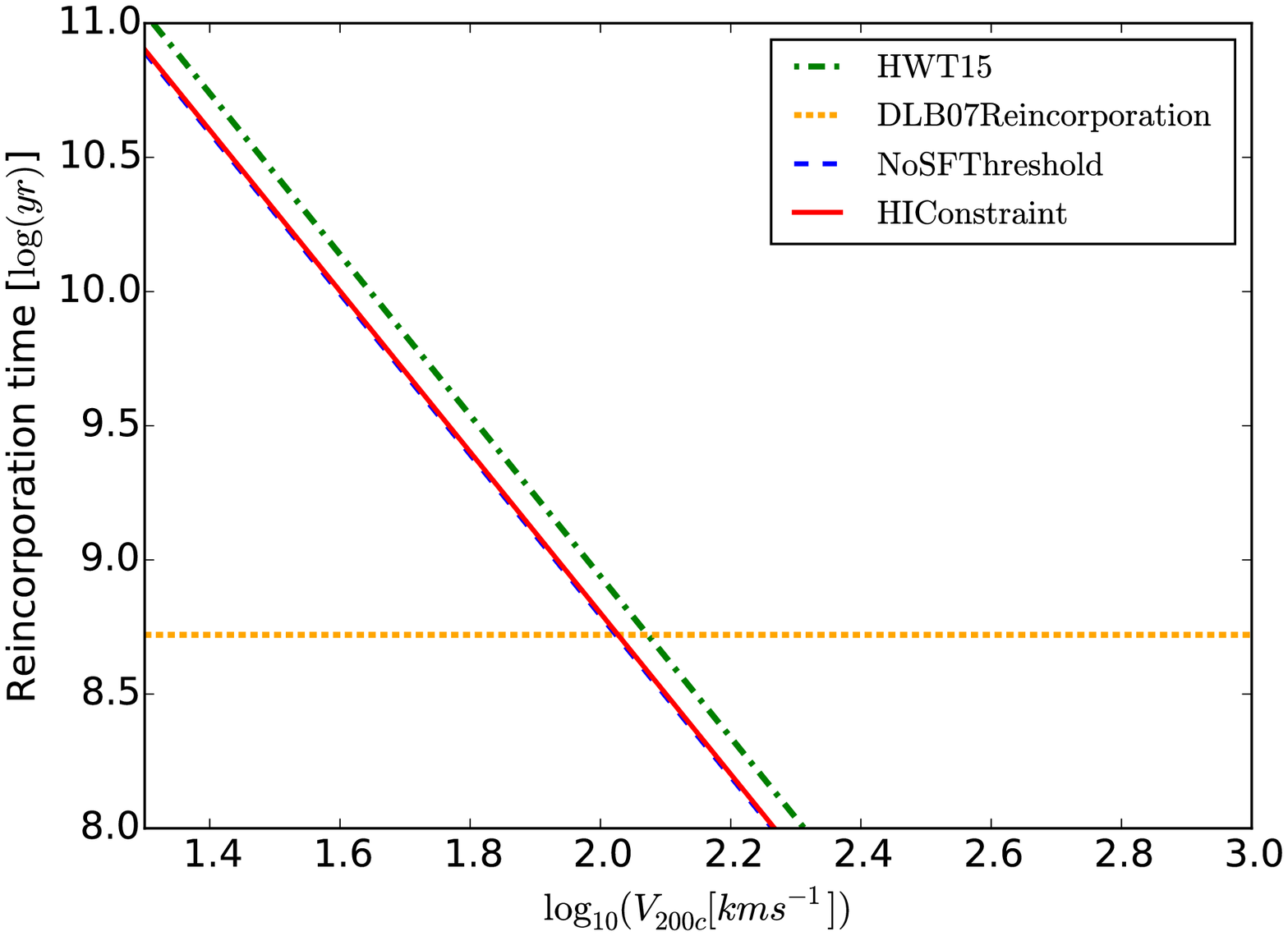}
\end{subfigure}

\caption{Supernova feedback parameters as functions of the halo velocity either
  maximum circular velocity, $V_{max}$ or virial velocity $V_{200c}$.  Top left
  is the SN Ejection Efficiency: the fraction of available SN energy for use
  in gas reheating and ejection.  Top right is the Mass-loading Efficiency, that
  controls how much cold gas is reheated.  Bottom left shows a derived quantity,
  the ratio of the mass of hot ejected gas to cold gas mass turned into stars.
  Finally, bottom right shows the Reincorporation Timescale for ejected gas.  In
  all plots the colours represent the same models as described above. All plots
  are at $z=0$.}
\label{fig:feedback}
\end{minipage}
\end{figure*}

The best fit parameters for our models are shown in \Tab{parameters}.  When
adding in the \HI\ mass function constraint into the HIConstraint and
NoSFThreshold models several parameters have changed significantly from those of
the original HWT15 model.  The biggest change is to the surface density
threshold for star formation, $M_\mathrm{crit,0}$, that we imposed.  As described in \Sec{sfLaw} we
have freed further the threshold parameter to allow it to become very low, or
have forced its removal entirely, to allow a reduction in the \HI content of
low-mass galaxies.  As compensation the star formation efficiency has decreased,
preventing the over production of stars in more massive systems.
The parameters controlling the feedback processes have changed slightly compared
to HWT15.  In \Fig{feedback} we plot the formulae that control feedback as a
function of virial velocity.  These formulae can be found in the supplementary
material of HWT15.

The top-left panel of \Fig{feedback} shows that the new models prefer a sharp
reduction in SN ejection efficiency above a halo circular speed of about
100\,km\,s$^{-1}$, dropping to just 10-20 per cent at higher masses.  This
allows more retention of gas in high-mass systems.  Slightly unexpectedly the
mass-loading factors, shown in the top-right panel of the figure, are lower than
for the fiducial HWT15 model, except for DLB07Reincorporation that requires large
mass-loading in dwarf galaxies to offset the rapid reincorporation (and
subsequent cooling) of ejected gas (bottom-right panel).  Unfortunately for that
model, the expenditure of energy to heat extra cold gas results in a decrease of
mass ejected in those dwarfs for a given amount of star-formation (lower-left
panel); elsewhere that ratio is similar for all models over all masses.
 
Finally we identify a shift in the best fit value of the threshold between 
minor and major mergers with respect to the revised version of HWT15.
That work found that there is some tension the value of this parameter required 
to match observations of the fraction of red galaxies and that required to match galaxy 
morphologies. HWT15 decided to fix $R_{\mathrm{merge}}$ at 0.1, slightly compromising 
colours at $z=2$ to better match observed morphologies at $z=0$.
For the purposes of this paper, the main effect of a major merger is to destroy disks,
turning \HI gas into stars or reheating it into the hot phase.  The threshold has increased from 
0.1 in the HWT15 model to 0.33-0.43 in the new models.  This sharp increase means many 
fewer mergers will be classed as major, allowing retention of more cold gas in massive 
galaxies.  However, major mergers are also an important mechanism for creation of 
elliptical galaxies and the cessation of star formation.  Their decrease contributes to the 
deficit of red galaxies we see in the lower panel of \Fig{redfrac}. 

\subsection{Star formation}
\label{sec:KSlaw}
\begin {figure}
\includegraphics[width=8.26cm]{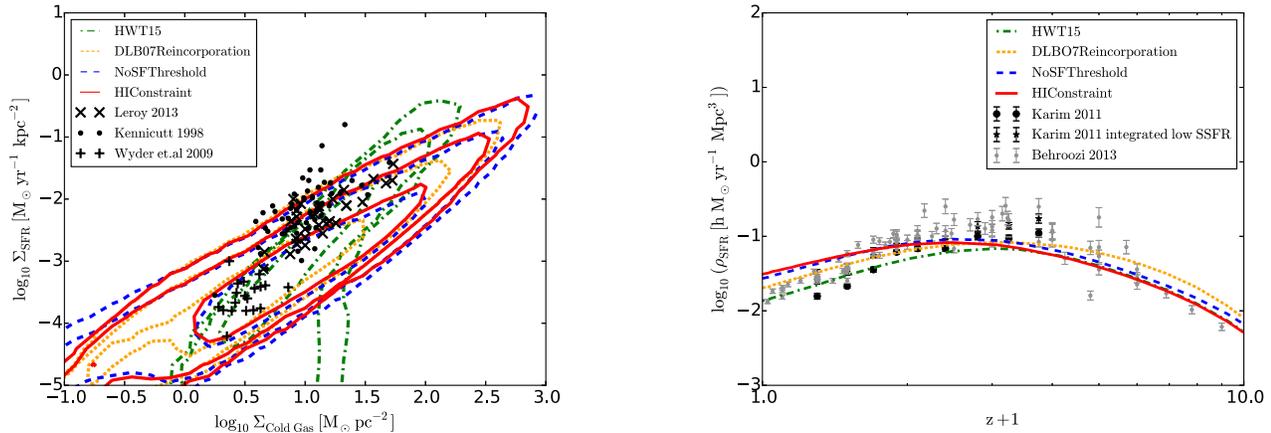}
\caption{Relationship between total gas surface density and the star formation
  rate surface density.The contours again enclose 68, 95 and 99 per cent of the
  data. The colours represent the same 4 models as previously. The black data
  points represent observed values from three different studies
  \protect\citep{Kennicutt1998,Leroy2013, Wyder2009}. }
\label{fig:KSLaw}
\end{figure}

\Fig{KSLaw} shows the effect that modifying our models has made to the
Kennicutt-Schmidt relation.  Both the observations and the model of HWT15 show a
break in the power law relation at low surface densities which is not reproduced
in the HIConstraint or NoSFThreshold models. The break arises
  naturally in the HWT15 from the finite threshold surface density for star
  formation.  Although not imposed as a constraint it seems to arise through a
  need to prevent galaxies being too blue at $z=2$.  Once we include the \HI
  mass function as a constraint, the break disappears because the improvement in
  that fit far outweighs the deterioration in the colours. We also see a
shallower slope which is similar to that observed between \HII surface density
and star formation rate \citep{Bigiel2008,Wyder2009}. This is perhaps an
indicator that we should form stars only out of the \HII component, although we
show in Appendix~A below that this does not, of itself, resolve the issues that
we see here.

\begin{figure}
\includegraphics[width=8.26cm]{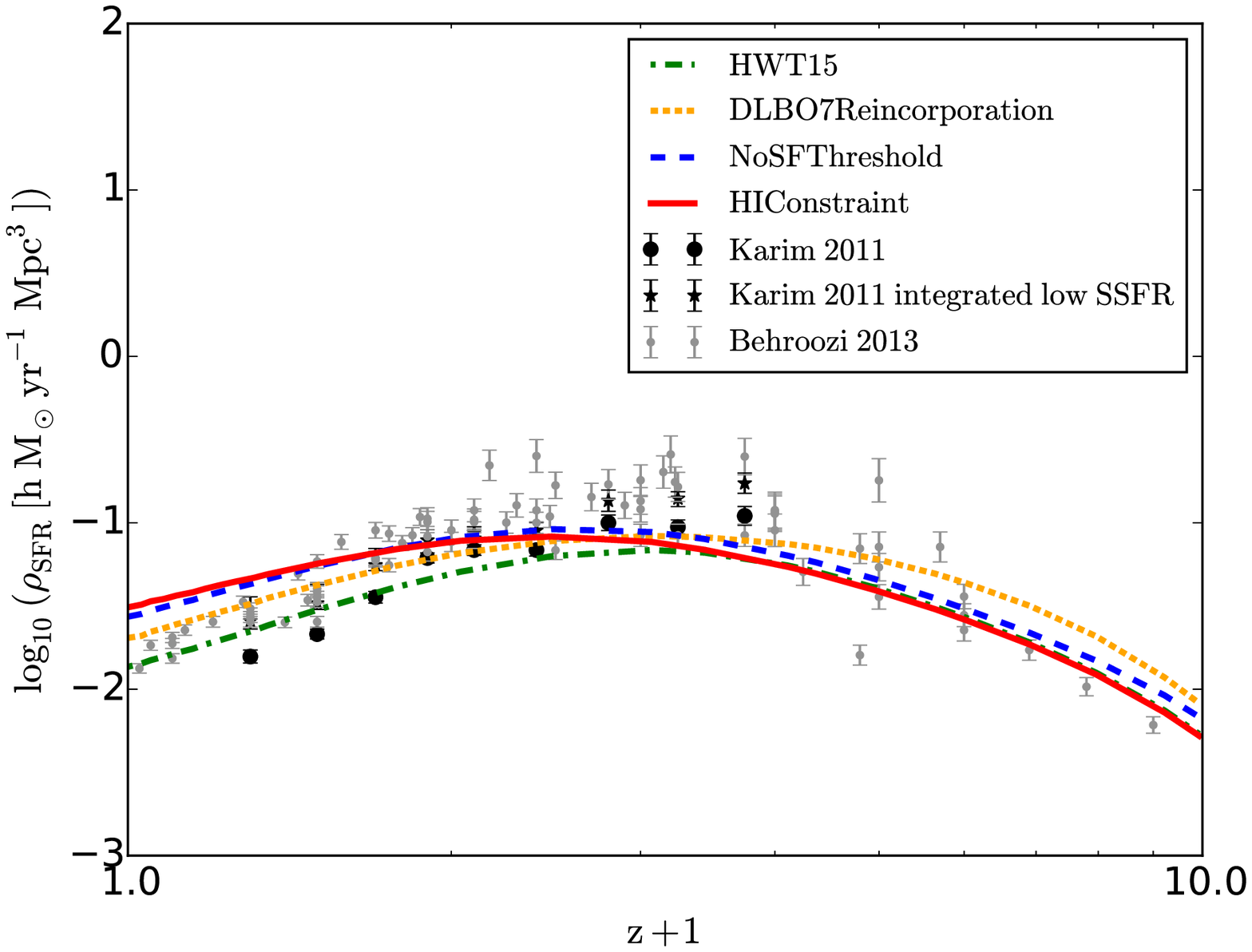}
\caption{The cosmic star formation rate density . The colours again represent
  the 4 models. These are compared to observations with the black data points
  from \protect\cite{Karim2011} and the grey from \protect\citet{Behroozi2013}}
\label{fig:sfrdens}
\end{figure}

In \Fig{sfrdens} we plot the star formation rate density(SFRD).  All semi-analytic models tend to produce SFRDs that evolve too weakly at low redshift and L-Galaxies is no exception. 
At $z=2$ all the models are very similar, while we start 
seeing more star formation in the new models at lower redshifts.
By $z=0$ there is significantly more star formation in the HIConstraint and NoSFThreshold models than in observations or HWT15.
This is likely contributing to the deficit of red galaxies seen
in \Fig{redfrac}.

A more detailed gas division model such as that used in \cite{Fu2010,Fu2012} could  
solve the problems presented in this section. \cite{Fu2010,Fu2012} analysed the impact of different 
star formation and gas division recipes with spatially-resolved discs producing a match to the 
observed \HI mass function (they still found an excess of dwarfs in the stellar mass function at $z=2$, but
this can likely be solved by the HWT15 gas reincorporation recipe). Spatially-resolved discs have not yet 
been implemented in the latest version of the Munich model. In order to try to understand the impact
of these modifications in our work we have implemented a simplified version of the Fu model. This is
described in \App{Appendix1} and goes some way to reconciling the Kennicutt-Schmidt relation with 
the \HI mass function. There is therefore some indication that a more realistic gas division along with 
adjustments to the star formation relation may be the solution.

\subsection{Comparison with other work} 
\label{sec:otherwork}

Lu14, who also use MCMC techniques to simultaneously fit the \HI mass function
and the $K$-band luminosity function, obtain much poor fits than we find and
claim that generic deficiencies of current SAMS are: (i) extreme mass-loading
factors are required in low-mass halos to expel the \HI; (ii) the outflow
requires more than 25\, per cent of the available supernova energy; and (iii)
the star-formation histories of Milky-Way sized halos are far too flat.

We do not require extreme mass-loading factors to achievement the 
agreement with observations presented in this paper. As shown in Fig. S2
of HWT15, the values we assume are comparable to current observational estimates.
On the other hand, we do require most of the supernova(SN) energy available to be used to 
power feedback. However, due to the uncertainties in the amount of energy produced 
by individual SN events we do not believe this rules out the models.

 In an attempt to understand the differences in our findings we have undertaken a run
using the reincorporation model of \cite{DeLucia2007} which more closely matches
that of Lu14.  We do not get such a good fit to the \HI mass function shown in
\Fig{himf}.  As can be seen in the top left panel of \Fig{feedback}, with the
DLB07 reincorporation recipe we find we require large mass loading factors in
low mass galaxies and still don't get a good fit for the \HI mass function. This 
could partially explain the differences between our results and those of Lu14.

\cite{Fu2010,Fu2012} integrate a model of gas division into a previous version of 
L-Galaxies, forming stars out of only the \HII component without using MCMC to constrain the parameters.
The model of gas division they use is more complex than that which we implement and the star formation 
recipe has no dependence on dynamical time. In addition, in regions where the molecular gas dominates 
the star formation goes as  $\Sigma_{\mathrm{SF}} \propto \Sigma_{\mathrm{H2}}$, while where atomic gas dominates $\Sigma_{\mathrm{SF}} \propto \Sigma_{\mathrm{gas}}^2$. Their work successfully reproduces the \HI mass function. However, as discussed in \Sec{KSlaw} they do not reproduce the low mass end of the stellar mass function as well.
Combining the work of Fu with HWT15 in future models of L-Galaxies could provide a solution to 
simultaneously producing the star forming properties and the \HI mass function. This is hinted at in \App{Appendix1}.

Similar work has been undertaken in the Galform model by \cite{Lagos2011b,Lagos2011a} using the same pressure gas division model as used in this work. The gas division was included self consistently with stars being formed out of the \HII component. They successfully reproduced the \HI mass function but did not reproduce the stellar mass functions as well. 
\cite{Popping2014,Somerville2015} also implement gas division in their semi-analytic model. They use several models of gas division and star formation and like \cite{Lagos2011b,Lagos2011a} they form stars from the \HII component. They successfully reproduce several \HI observations of galaxies. Their \HI mass function exhibits a slight excess at low masses but fits well at the high mass end.

\section{Conclusions}
\label{sec:conclusions}

In this paper we have added the \HI mass function as an observational constraint
to the L-Galaxies semi-anlaytic model of \cite{HWT15}.  Using MCMC techniques we
re-constrain the model parameters in order to best fit this extra observation at
$z=0$ in addition to the stellar mass function and galaxy colours at $z=0$ and
$z=2$.  The cold gas content of the model galaxies are divided in post
processing into the \HI and \HII components using the gas division model of
\cite{Blitz2006} and the approximation to this from \cite{Obreschkow2009}.

From this work we conclude :
\begin{enumerate}
\item Using the $z=0$ \HI mass function as an extra constraint we obtain a good
  fit to this in addition to the stellar mass function and red fraction at $z=0$
  and $z=2$. 

\item The most important parameter change is the reduction of the star formation
  gas surface density threshold. This has been greatly reduced or even
  removed. This was required to remove the excess of \HI gas seen in low mass
  galaxies in HWT15.  As compensation, the star formation efficiency has
  decreased, preventing the over production of stars in more massive systems.

\item The feedback parameters have also changed. The retuned model
  favours a sharp reduction in the SN ejection efficiency above a halo circular
  speed of 100\,km\,s$^{-1}$ to much lower efficiencies compared to HWT15. 
  The required mass loading factors are also reduced slightly compared to HWT15.

\item The model has a worse fit to the star formation properties shown in the
  Kennicutt-Schmidt relation and the cosmic star formation rate density at low
  redshifts. We see too much star formation $z=0$, mostly in the low mass
  galaxies.  This suggests that we either incorporate and cool too much
    gas, or that we underestimate the expulsion of gas via winds and stripping.
    However, since our red fractions roughly agree with observations,
    any changes must only reduce the star formation efficiency and not halt it
    completely.

\item We use the reincorporation model of DLB07 to compare our model with that
  of Lu14. We alleviate some but not all of the problems identified by Lu14 through using
  an alternative reincorporation recipe. It is likely that a detailed model gas division 
  and subsequent star formation will be required
  to match the observations.
\end{enumerate} 

Using a more detailed model of cold gas division and a change to the star formation recipe, such as those used in \cite{Fu2010,Fu2012,Fu2013}, we expect to improve on the problems with simultaneously matching both the star formation properties and the observed \HI mass function. In \App{Appendix1} we show a simplistic model in which we use the approximation for gas division given in \Sec{HImodel} and then form stars only out of the molecular gas component. While the resulting \HI mass function is not as good a fit as our HIConstraint model it is a significant improvement on the original HWT15 fit. Likewise for the Kennicutt-Schmidt relation the model shown in \App{Appendix1} is an improvement on the HIConstraint model show in \Fig{KSLaw}. 

In summary, the cold gas mass function provides a useful constraint on galaxy
formation models that poses challenges to the current paradigm.  It is difficult
to lower the \HI mass function in low-mass galaxies without violating the
Kennicutt-Schmidt star formation law and having too much star-formation in dwarf
galaxies in the current-day Universe. It is likely that a detailed model of the cold gas in the \HI and \HII
  components and subsequent star formation is required to resolve the issue.

%% file: appendix1.tex
\section{Star formation from molecular gas}
\label{app:Appendix1}
\begin {figure}
\includegraphics[width=8.26cm]{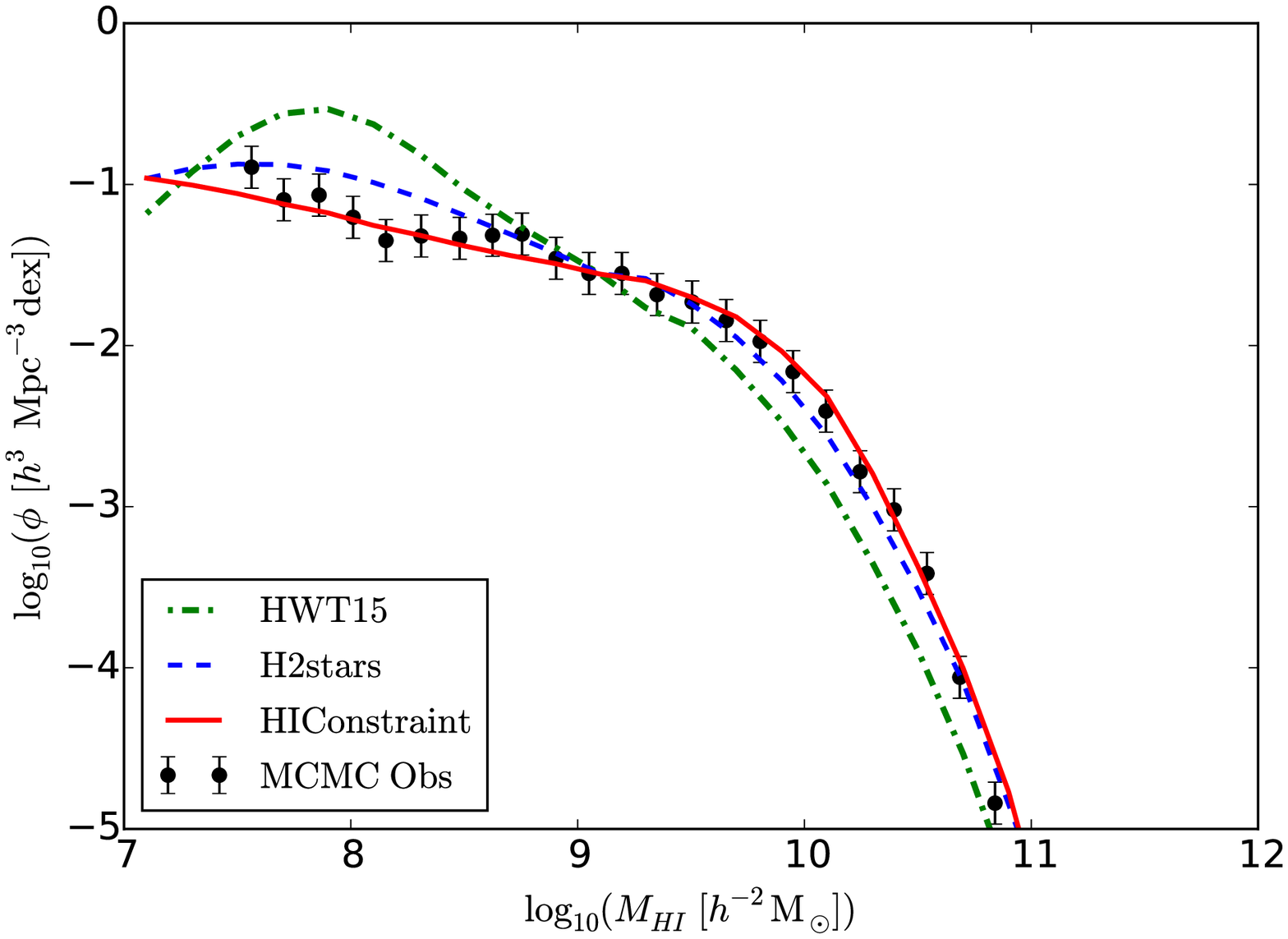}
\caption{The \HI mass function. The red and green lines are as in previous
  figures; the blue line uses the gas division approximation to form stars out
  of only \HII gas.}
\label{fig:HI_H2}
\end{figure}

\begin {figure}
\includegraphics[width=8.26cm]{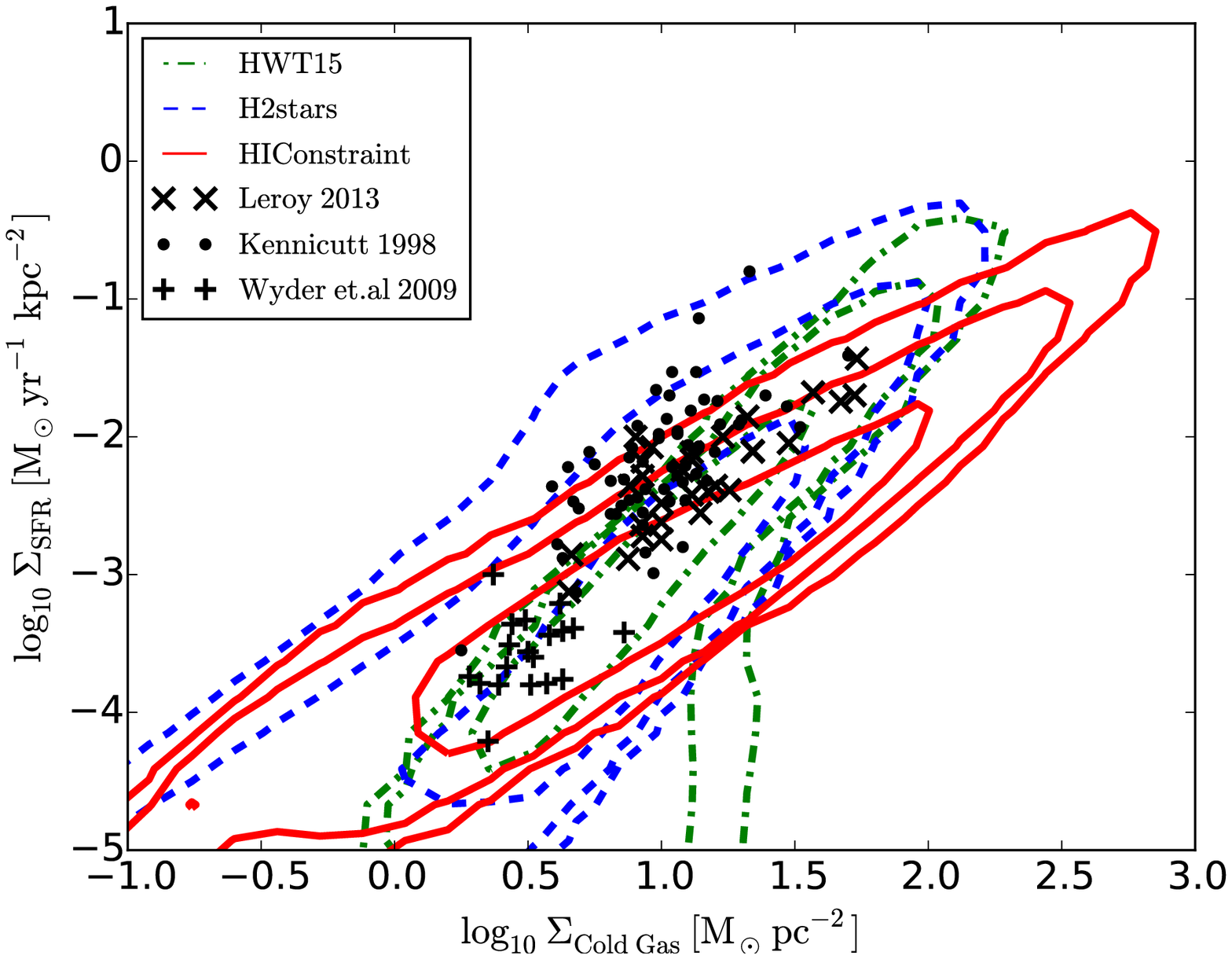}
\caption{Relationship between total gas surface density and the star formation
  rate.  The contours again enclose 68, 95 and 99 per cent of the data. The red and
  green are as in previous figures and the blue uses the gas division
  approximation to form stars out of only \HII gas.  The black data points
  represent observed values from three different studies
  \protect\citep{Kennicutt1998,Leroy2013, Wyder2009}. }
\label{fig:KS-H2}
\end{figure}

We have investigated the effect of using the approximation given in \Eq{approx}
in order to form stars out of only the \HII component of the cold gas.  We
modify \Eq{SFLaw} so that the gas mass is that of just the \HII component and
there is no longer any gas density threshold. The resulting \HI mass function
and Kennicutt-Schmidt relation are shown in \Fig{HI_H2} and \Fig{KS-H2},
respectively.  In the \HI mass function we see a slight excess of galaxies with low HI
masses, significantly better than the original HWT15 but slightly worse than our best fit
HIConstraint model.  The new model roughly fits the slope of KS relation, although it
might not have a sharp enough break at low masses. We conclude that the formation 
of stars out of only the \HII component gives an interesting compromising in the comparison
between model and observations for the \HI mass function and KS relation. A detailed
model of \HII conversion and subsequent star formation might correct the excessive 
cold gas in the lowest mass galaxies.

